\documentclass[aps,prd,reprint,superscriptaddress]{revtex4-1}
\usepackage[utf8]{inputenc}
\usepackage{amsmath}
\usepackage{amssymb}
\usepackage{graphicx}
\usepackage{color}

\usepackage{psfrag}

\usepackage[normalem]{ulem}

\newcommand{\MM}{{\cal M}}

\newcommand{\SUIII}{\text{SU}(3)}

\begin{document}

\title{Lattice QCD on Non-Orientable Manifolds}

\author{Simon Mages}
\email{simon-wolfgang.mages@ur.de}
\affiliation{J\"ulich Supercomputing Center, J\"ulich D-52425, Germany}
\affiliation{University of Regensburg, Regensburg D-93053, Germany}

\author{B\'alint C. T\'oth}
\email{tothbalint@szofi.elte.hu}
\affiliation{University of Wuppertal, Department of Physics, Wuppertal D-42097, Germany}

\author{Szabolcs Bors\'anyi}
\affiliation{University of Wuppertal, Department of Physics, Wuppertal D-42097, Germany}

\author{Zolt\'an Fodor}
\affiliation{J\"ulich Supercomputing Center, J\"ulich D-52425, Germany}
\affiliation{University of Wuppertal, Department of Physics, Wuppertal D-42097, Germany}
\affiliation{E\"otv\"os University, Budapest 1117, Hungary}

\author{S\'andor D. Katz}
\affiliation{E\"otv\"os University, Budapest 1117, Hungary}

\author{K\'alm\'an K. Szab\'o}
\email{szaboka@general.elte.hu}
\affiliation{J\"ulich Supercomputing Center, J\"ulich D-52425, Germany}
\affiliation{University of Wuppertal, Department of Physics, Wuppertal D-42097, Germany}

\date{10 April 2017}

\begin{abstract}

    A common problem in lattice QCD simulations on the torus is the extremely
    long autocorrelation time of the topological charge, when one approaches
    the continuum limit. The reason is the suppressed tunneling between
    topological sectors. The problem can be circumvented by replacing the torus
    with a different manifold, so that the connectivity of the configuration
    space is changed. This can be achieved by using open boundary conditions on the
    fields, as proposed earlier. It has the side effect of
    breaking translational invariance strongly. Here we propose to use a
    non-orientable manifold, and show how to define and simulate lattice QCD on
    it. We demonstrate in quenched simulations that this leads to a drastic
    reduction of the autocorrelation time. A feature of the new proposal is,
    that translational invariance is preserved up to exponentially small
    corrections. A Dirac-fermion on a non-orientable manifold poses a challenge
    to numerical simulations: the fermion determinant becomes complex. We
    propose two approaches to circumvent this problem.

\end{abstract}

\pacs{}

\maketitle

\section*{Introduction}

Lattice regularization is a powerful method to carry out calculations in a
quantum field theory. It provides a well-defined, systematically improvable
framework, that also works in the non-perturbative regimes of the theory. There
is currently a lot of activity to calculate various observables in Quantum
Chromodynamics (QCD) by carrying out numerical computations on lattices.
Such activities are an important cornerstone in the search for
physics beyond the Standard Model. In many cases lattice results with a
precision beyond the \% level are needed to fully exploit the discovery
potential of these searches. To reach such a precision it is important to have
a reliable error estimation.

One type of error comes from the autocorrelation in the Monte Carlo time series of
the numerical simulations. In principle
one has to run the simulation several times longer than the largest
autocorrelation time in the system.  Typically the slowest modes correspond
to observables which are related to the topology of the field space, like the
topological charge:
\begin{align}
	Q=\int_{\MM} \text{d}^4 x\ q(x),\label{eq:global_charge}
\end{align}
where $q(x)=\frac{1}{32\pi^2}\epsilon_{\mu\nu\rho\sigma}\mathrm{tr}F_{\mu\nu}F_{\rho\sigma}$ is the topological charge density.
In practice the space-time manifold $\MM$ is chosen to be the torus, where
periodic/anti-periodic boundary conditions are imposed on the fields. On the torus $Q$ is
quantized and the field space splits into disconnected sectors labeled by integer  \cite{vanBaal:1982ag} values of $Q$.
The advantage of the torus is translational invariance, and as a consequence the results
have small finite volume corrections. The disadvantage is that conventional
simulation algorithms have severe difficulties changing the topological properties of the
field configurations, and it gets worse with decreasing lattice spacing. The
autocorrelation time of the topological charge was found to increase with the
sixth power of the inverse lattice spacing in actual simulations
\cite{Schaefer:2010hu}. Rare tunneling events make the extraction of the
topological susceptibility challenging. For a recent proposal see \cite{Brower:2014bqa}.
Besides the susceptibility, the accurate computation of observables that
correlate strongly with topology is also challenging.

The slow modes can be removed from the theory by changing the topology of the
manifold $\MM$. The authors of \cite{Luscher:2011kk} proposed to introduce an
open boundary in one of the directions.  This change in the topology of
space-time also changes the topology of the gauge field configuration space, which becomes
connected and $Q$ is not restricted to an integer value any more.  This
eliminates the slow modes from the theory.  Indeed in \cite{Luscher:2011kk} a
drastic reduction of the autocorrelation time of the topological charge was
observed. A disadvantage of this approach is the lack of the translational
invariance in the open direction, which introduces boundary effects and
decreases the effectively available space-time volume. 
There are several recent studies with open boundary conditions,
which address the systematics of the method \cite{Luscher:2012av,Bruno:2014lra,Amato:2015ipe,Lucini:2015wvo}.

Another possibility to circumvent large autocorrelation times is to restrict
the simulation to a single topological sector.  This removes the slow modes
corresponding to the changes between the sectors.  To extract the topological
susceptibility from fixed sector simulations see the Refs.
\cite{Aoki:2007ka,Bietenholz:2015rsa,Bautista:2015yza}. However fixing the
topology introduces finite volume effects, that are proportional to the inverse
of the volume \cite{Brower:2003yx}. Additionally it is not clear if fixing the
topological sector is compatible with ergodicity, and if not, it is an open
question how the observables are affected. Let us also mention, that there are
recent ans\"atze, which increase tunneling between different topological
sectors by modifying the original theory and applying a reweighting correction
in the observables \cite{Kitano:2015fla,Laio:2015era}. Here the efficiency of
the reweighting might limit the applicability of these methods. Also it was
proposed to modify the action by dislocation enhancing determinant ratios
(DEDR) to improve the tunneling \cite{McGlynn:2013ava}. Finally, 
there is a proposal to use a multi-level thermalization scheme to sample
the topology better on a fine level \cite{Endres:2015yca,Detmold:2016rnh}. The
validity of this strategy might currently be limited by the sampling of the
topology on the coarse level which is prolonged to and frozen on the fine
level.

\section*{Simulating on non-orientable manifolds}

\begin{figure}
    \centering
	\includegraphics{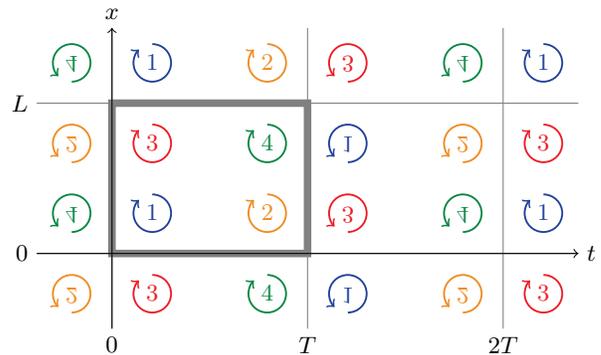}
	
	\caption{
	    Illustration of the $P$-periodic boundary condition in the $t$ direction. 
	    The orientation changes when the boundary
	    is crossed as indicated by the arrows.
	    In the
	    $x$ direction we have the usual periodicity.
	    The base manifold is the rectangle with bold lines. 
	}
	\label{fi:sketch}
\end{figure}

In this paper we propose a solution that preserves the translational symmetry up to
exponentially small corrections and alleviates the problems with frozen topological charge:
simulate the theory on a non-orientable manifold. To construct such a manifold
let us start from a $L^3\times T$ torus with spatial size $L$ and temporal size
$T$.  Now we replace the periodic boundary condition in the temporal direction
by a ``$P$-periodic'' boundary condition: the fields are parity transformed
across the boundary.  With this boundary we get a manifold which locally looks
like the torus, but is different globally.  The boundary condition can be
imagined as an infinite manifold, which is split into blocks of size
$L^3\times T$ with one of the blocks being the original manifold.  In the spatial
directions the blocks are replications of the original manifold, whereas the
neighbor in the temporal direction is obtained by parity transforming the
original block. This is illustrated for two dimensions, one spatial and one
temporal, in Figure \ref{fi:sketch}, in which case the base manifold is just the Klein-bottle. If
we replaced the periodic spatial boundary condition with open, the manifold would be the M\"obius-strip.

The straightforward way to define a global topological charge on a
non-orientable manifold via the integral of a local topological charge
density over the complete manifold, does not work: on a non-orientable manifold
there is no global volume form to define integration. There is however a global
volume element and one can define the integration of scalar densities like the
action density. But one can not use this to also define the integration of a pseudo-scalar
density over the full non-orientable manifold. As a workaround we define a total
charge $Q_m$ on a maximal oriented submanifold
\begin{align}
	Q_m=\int_0^T \text{d}t\int \text{d}^3x q(x).
\end{align}
Here the integration is performed over the manifold with a cut at $t=0$,
which makes the volume form well defined. This definition is the same as
on open boundary lattices, but with an ad-hoc introduced cut without
physical meaning. In the following we drop the index $_m$ and define
$Q:=Q_m$.

Then it is simple to show that on a non-orientable manifold this charge, defined
as an integral of $q(x)$ over the base manifold, is not quantized. Under a
parity transformation $P$ the topological charge transforms to its negative. Therefore
applying a continuous translation on the gauge field in the time direction
changes the charge continuously. After a translation of $T$ we get the same
charge value, that we started with, but with an opposite sign. Since the charge
varies continuously during the translation, it cannot possibly be quantized.
Let us note, that the charge over the double cover, i.e. defined as an
integral running from $0$ to $2T$ in time, is zero.
However our setup is different from fixing the topology of a $L^3\times 2T$
lattice. 
The constraint in a fixed topology simulation is
non-local, and leads to finite volume effects proportional to the inverse
volume \cite{Brower:2003yx}. In contrast the $P$-periodic boundary condition gives a local quantum field
theory by construction.

$P$ invariant quantities, like the gauge and fermion actions in QCD, are
invariant under a translation in the $t$-direction. In contrast $P$
non-invariant observables change after such a translation, as we have seen
already in the case of $q(x)$. To avoid large finite volume effects in $P$
non-invariant observables some care is necessary, see later. The translations
in the $x,y,z$ directions and Euclidean rotations are not exact symmetries with
the $P$-boundary condition. The violations of these symmetries originate from Feynman diagrams,
where a particle propagates around the $t$-direction. Therefore they have to be
exponentially suppressed with $T$ times the massgap of the theory.

The $P$-periodic boundary condition is similar to the $C$-periodic boundary
condition of Refs. \cite{Wiese:1991ku,Kronfeld:1990qu,Lucini:2015hfa}, where a
charge conjugation is performed when the boundary is crossed. They lead to
processes, which change the total electric charge: charge fluctuations when
propagating through the $C$-periodic boundary change their sign and thereby the
total electric charge.  Analogously our $P$-periodic boundary condition leads
to changes in the total topological charge. Also, just as in the
$C$-periodic case, the ultraviolet structure of the theory is not affected by
the boundary condition and the same renormalization applies as in infinite
volume \cite{Lucini:2015hfa}.

The implementation of the parity transformation on parallel computers can be
cumbersome. Therefore we choose another transformation, which serves the
purpose equally well: the lattice points shall be reflected through the
$x=0$ hyperplane. Since this transformation is a product of $P$ and a rotation
by 180$^{\circ}$ around the $x$-axis, it also changes the orientation
and defines a non-orientable manifold.  For simplicity
we use the name ``$P$-periodic'' boundary condition for this setup in the
rest of the paper.

\section*{Gauge fields}
\label{sec:gauge}

To demonstrate the viability of our proposal we performed numerical
simulations in pure $\SUIII$ gauge theory.
The prescription for the $P$-boundary is:
\begin{align}
    \begin{split}
    U_{x}(x,y,z,t+T)&= U^\dag_{x}(L-x-1,y,z,t),\\
    U_{i}(x,y,z,t+T)&= U_{i}(L-x,y,z,t)
    \end{split}
    \label{eq:pper}
\end{align}
for $i=y,z,t$. In the other three directions we keep the usual periodic boundary condition.
We use the
tree level Symanzik improved
action \cite{Luscher:1985zq} and lattices of a fixed physical size of $L=T\sim2.27/T_c$. One update
sweep consists of four overrelaxation and one heatbath steps \cite{Cabibbo:1982zn,Kennedy:1985nu,Creutz:1987xi}. There
is practically no overhead on the simulation time coming from the $P$-boundary condition.
We chose
five different lattice spacings. The lattice size, gauge coupling, $w_0$ scale
\cite{Borsanyi:2012zs}, the lattice spacing, and the number of update sweeps are
given in Table \ref{tab:LatticeParameters}. The lattice spacings shown are
obtained from the conversion $a=0.167\ \mathrm{fm}/w_0$ \cite{Sommer:2014mea}. For comparison
we simulate three streams at every set of parameters: one with periodic, one
with open, and one with $P$-periodic boundaries.  Our main observables are the
topological charge $Q$ and time slice averages of the topological charge and
action densities $Q(t)$ and $E(t)$ as defined in \cite{Luscher:2011kk}.  All
are evaluated along the Wilson flow \cite{Luscher:2010iy} at a flow time of $w_0^2$.

\begin{table}[h]
    \centering
    \begin{tabular}{c|c|c|c|c}
	$L$ & $\beta$ & $w_0$ & $a$[fm] & $n_{sweep}$ \\
	\hline
	16 & 4.42466 & 1.79 & 0.093 & $ 2\times4001$ \\
	20 & 4.57857 & 2.24 & 0.075 & $ 3\times4001$ \\
	24 & 4.70965 & 2.65 & 0.063 & $ 4\times4001$ \\
	32 & 4.92555 & 3.43 & 0.049 & $10\times4001$ \\
	40 & 5.1     & 4.13 & 0.040 & $19\times4001$
    \end{tabular}
    \caption{Lattice Parameters}
    \label{tab:LatticeParameters}
\end{table}

\begin{figure}
    \centering
    \includegraphics{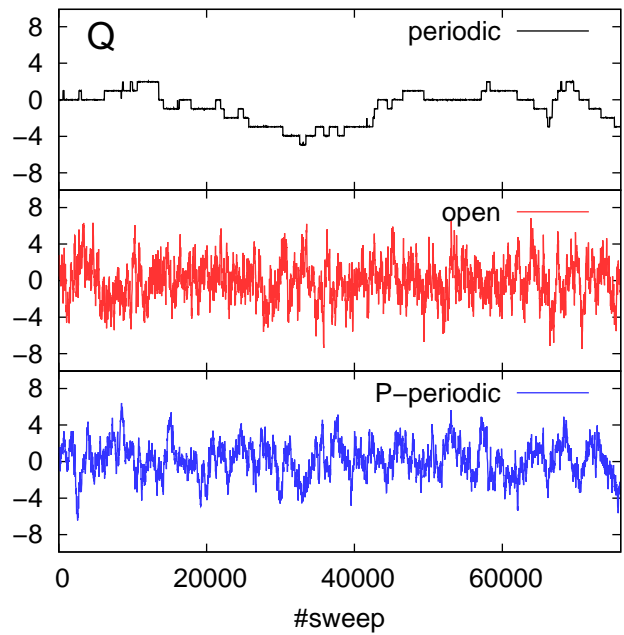}

    \caption{History of the topological charge $Q$
    at $\beta=5.1$, the corresponding lattice spacing is $0.040$ fm.\label{fig:history}}

\end{figure}

Figure \ref{fig:history} shows the simulation time history of the topological
charge for the finest lattice spacing. It already shows a drastic reduction of the
autocorrelation of $Q$.

\begin{figure}
    \centering
    \includegraphics{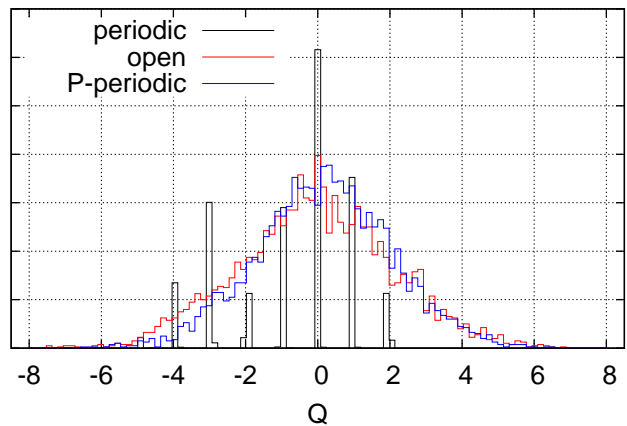}

    \caption{Histogram of the topological charge $Q$
    at $\beta=5.1$, the corresponding lattice spacing is $0.040$ fm.\label{fig:histogram}}

\end{figure}

The discrete nature of the topological charge in the
periodic case can be seen best in a histogram of the topological charge,
which is given in Figure \ref{fig:histogram} together with the histograms
of the charges with the other two boundary conditions, which show no sign of
discretization.

\begin{figure}
    \centering
    \includegraphics{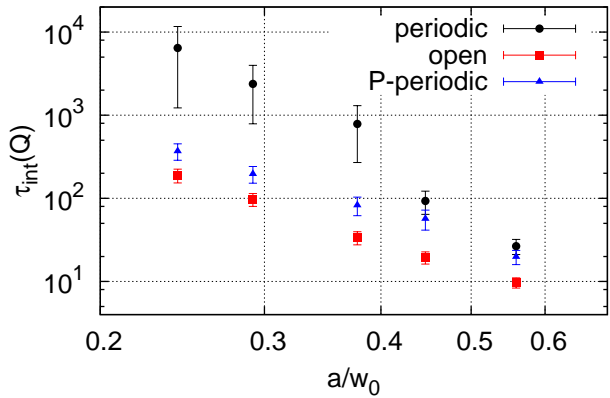}

    \caption{Integrated autocorrelation time of the topological charge
    as a function of the lattice spacing. \label{fig:tint}}

\end{figure}

The lattice spacing dependence of the integrated autocorrelation time of the
topological charge $\tau_{int}(Q)$ is given in Figure \ref{fig:tint}. Again one
can clearly see that both open and $P$-periodic boundaries give a strong
reduction of $\tau_{int}(Q)$ compared to the periodic case. Though
$P$-periodic simulations have somewhat larger autocorrelation times than the open simulations, they seem
to scale with a similar power of the lattice spacing. 

In \cite{McGlynn:2014bxa}
a simple model was set up to describe the scaling of $\tau_{int}(Q)$
with the lattice spacing. 
There the autocorrelation function of
the topological charge is investigated: $\Gamma(t,\tau)= \langle Q(t)_\tau
Q(0)_0 \rangle$, where $Q(t)_\tau$ is the time-slice topological charge after
$\tau$ update sweeps.  A diffusion equation for $\Gamma$ is set up, whose
solutions describe numerical data from periodic and open lattices quite
accurately.  The extremely large autocorrelation times of the topological
charge on periodic lattices are attributed to the presence of zero time-like
momentum modes in $\Gamma$. They arise due to the periodicity of $\Gamma$ in
the $t$-variable: $\Gamma(t+T,\tau)=\Gamma(t,\tau)$, which is the consequence
of the periodic boundary $Q(t+T)=Q(t)$. These zero modes are shown to be
eliminated by changing the boundary condition from periodic to open
\cite{McGlynn:2014bxa}. Repeating the calculation for the $P$-boundary,
we see that the same elimination occurs.
The zero modes are absent since, as explained before, the charge is
anti-periodic $Q(t+T)=-Q(t)$, and therefore $\Gamma$ is also anti-periodic:
$\Gamma(t+T,\tau)=-\Gamma(t,\tau)$. Without these zero-modes the autocorrelation
time of $Q$ scales similarly to that of the local observables. Let us note,
that the diffusion and local tunneling parameters of this model do not depend on
the boundary condition.
Our findings are in agreement with the qualitative picture taken from this
model calculation.

\begin{figure}
    \centering
\includegraphics{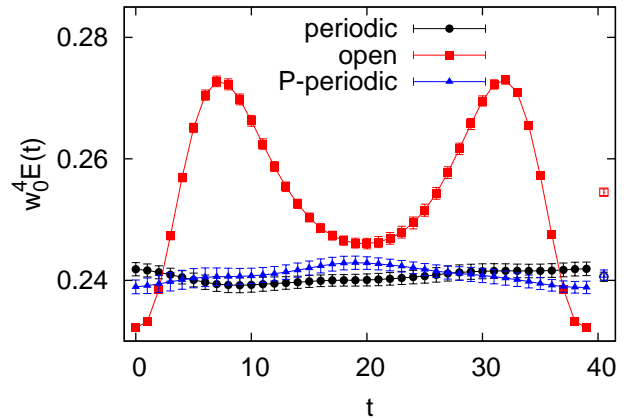}

\caption{Time slice averaged action density at $\beta=5.1$, lattice spacing $a=0.040$ fm, and box size $1.6$ fm. Filled symbols give the per time slice results, while open symbols give the result for the full volume.\label{fig:slices}}

\end{figure}

\begin{figure}
    \centering
    \includegraphics{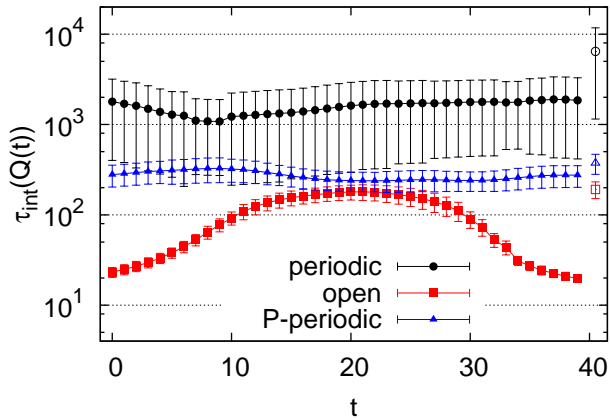}

\caption{Integrated autocorrelation time of the topological charge on a time slice at $\beta=5.1$, lattice spacing $a=0.040$ fm, and box size $1.6$ fm. Filled symbols give the per time slice results, while open symbols give the result for the full volume.\label{fig:tintslices}}

\end{figure}

Now we come to the most important feature of our proposal. In contrast to the
open boundary condition the $P$-periodic boundary is translationally invariant in the $t$-direction by
construction. This is shown in Figures \ref{fig:slices} and \ref{fig:tintslices}. Figure \ref{fig:slices} gives the
time-slice averaged action $E(t)$ as the function of time.  For the usual and
$P$-periodic boundaries $E(t)$ are practically constant and agree well with
each other, reflecting translational symmetry and the absence of large finite
volume effects.  The open boundary result deviates from them significantly. Although it
is expected to approach the periodic value in the middle of the lattice, here
the time extent was not large enough to reach it.

Similar effects can also be seen in Figure \ref{fig:tintslices}, which
gives the integrated autocorrelation time of the topological charge on a
time slice as a function of time. Again the results for periodic and
$P$-periodic boundaries are independent of time, while the result for open
boundaries has a dependence on the distance from the boundary. Additionally
this plot shows that compared to the full volume results the integrated
autocorrelation time of the topological charge on a single time slice improves
for all choices of the boundary. In the periodic case, however, the result for
a single time slice is larger than the result for open and
$P$-periodic boundaries. The reasoning of the subvolume method \cite{Brower:2014bqa}
would suggest that the autocorrelation time on the smallest subvolume -- a
single time slice -- would be the same independent of the boundary conditions.
In the diffusion model one can see why this might be not the case: there is a zero
mode contributing to the autocorrelation time only in the periodic case, but
not in the open or $P$-periodic case, which increases the autocorrelation.

\begin{figure}
    \centering
    \includegraphics{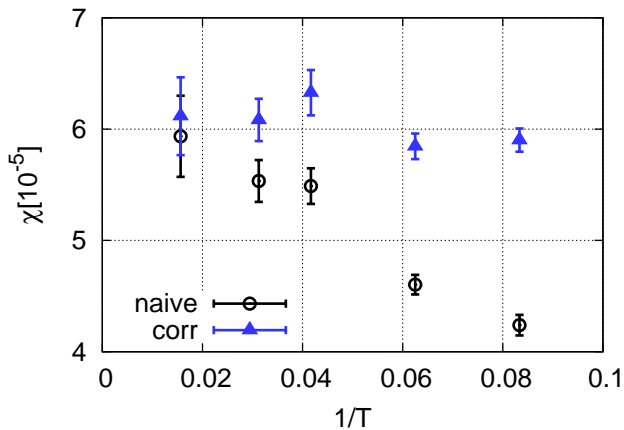}
    \caption{
	\label{fi:ldep} Finite size dependence of different topological susceptibility definitions for $P$-periodic boundary conditions
	at $\beta=4.42466$, lattice spacing $a=0.093$ fm.
    } 
\end{figure}

The reason for $Q$ being not quantized with $P$-periodic boundary condition, is
the lack of translational invariance for $Q$ as described earlier. This demands
some additional care, when calculating an observable like the topological
susceptibility. The appropriate procedure uses the definition via the
topological charge density correlator, where we place the origin of the correlator
to timeslice $T/2$:
\begin{align}
    \label{eq:chi}
    \chi= \int_{0}^{T} \text{d}t\ \int \text{d}^3x\ \left\langle q(\vec{x},t) q(\vec{x},T/2) \right\rangle.
\end{align}
It has exponentially small finite volume corrections as $T$ is increased, since
when $t$ in the integral
is close to $0$ or $T$, the correlator is exponentially small.
Also this definition is symmetric in $t$ and leads to a maximal cancellation
of the negative tail of the correlator with its positive core.
To illustrate this we generated
dedicated ensembles at $\beta=4.42466$ with size $16^3\times T$ with
$T=12,\dots 64$. The results show no significant finite volume effects, as
shown in Figure \ref{fi:ldep} with label 'corr'. 
The more commonly used definition for the susceptibility is $\langle Q^2\rangle/(L^3T)$.
For periodic boundary condition the two definitions are equivalent due to the translational
invariance. In the $P$-periodic case the latter definition gives an observable,
that has finite size effects proportional to $1/T$,  
see the points with label 'naive' in Figure \ref{fi:ldep}.
Let us remark, that a similar prescription as in Equation \ref{eq:chi} can also improve
the finite volume correction of the standard subvolume method in periodic simulations.

\section*{Fermions}

While non-orientability can be implemented on the gauge fields right away, it
becomes non-trivial for fermions. As the boundary mixes the handedness it is
not possible to define Weyl-fermions on a non-orientable manifold. The
definition of Dirac-fermions is straightforward \cite{Grinstein:1987vn}, the fields
undergo an $x$-axis reflection on the boundary. However we encounter a serious
obstacle: the Dirac-matrix entering in the $x$-axis reflection ($\gamma_5\gamma_x$) spoils
the $\gamma_5$-Hermiticity of the Dirac operator and leads to a complex
fermion path integral.
Numerical simulations with standard algorithms are not possible in this case.
Here we show two workarounds.

The first solution is to combine the $x$-reflection with a charge conjugation.
Since the topological charge is not only $P$-odd but also $CP$-odd, this combination
has the same advantageous effect as the $P$-periodic boundary condition.
Instead of the usual $\psi$, $\overline{\psi}$ fields, it turns
out more 
convenient to work with the eigenstates of charge conjugation. These
are
the spinor fields $\eta_a$, $a=1,2$, defined in \cite{Lucini:2015hfa}.
In both bases we can write down the one flavor fermion action in infinite volume:
\begin{align}
    S_f= \overline{\psi}D[U]\psi = -\frac{1}{2} \eta_a C \hat{D}[U]_{ab} \eta_b,\label{eq:major}
\end{align}
where $D$ is some possibly massive lattice Dirac-operator
using the usual symmetric expression for the time derivatives, and
$C$ is the charge conjugation matrix.
The hatted Dirac-operator $\hat{D}[U]$ can be written with the
original Dirac operator as
$\hat{D}[U]_{ab}= D[\mathrm{Re}U\cdot\delta_{ab} -i\mathrm{Im}U\cdot \rho_{2,ab}]$ with
the Pauli-matrices $\rho_i$ acting on the $a$-index of the $\eta_a$ fields.
As the map $U \mapsto \mathrm{Re}U\cdot\delta_{ab} -i\mathrm{Im}U\cdot \rho_{2,ab}$
defines a representation of $\SUIII$ equivalent to {\boldmath$3 \oplus 3^\ast$},
Equation \ref{eq:major} is valid not only for Dirac operators which are linear in the links
but also for operators which are linear in products of links, such as the clover improved 
Wilson Dirac operator. Equation \ref{eq:major} is even valid for some more general cases,
like the overlap operator.
For the readers convenience we collect
the symmetries of $S_f$ in this somewhat uncommon representation
in the Appendix.
Now we introduce the ``$CP$-boundary'' condition:
\begin{align}
    \eta(x,y,z,t+T) &= -\gamma_5\gamma_x\rho_2\rho_3\ \eta(L-x,y,z,t),
\end{align}
where we used the $x$-reflection and charge conjugation operators defined in the Appendix.
The gauge fields also have to undergo
a $CP$-transformation, this means
additional complex conjugations in Equation (\ref{eq:pper}).
The resulting
Dirac-operator is $\gamma_5\rho_2$-Hermitian
\begin{align}
    \hat{D}^\dagger= \gamma_5\rho_2 \hat{D} \gamma_5\rho_2,
\end{align}
since the matrix $\gamma_5\rho_2$ commutes with the boundary condition
\cite{Metlitski:2015yqa}. Therefore the path integral over the $\eta$ fields
gives a real Pfaffian. In the massless case and in the continuum $\hat{D}$ also satisfies
\begin{align}
    \{\hat{D},\gamma_5\rho_2\}= 0 \quad\quad\text{for}\quad m=0.
\end{align}
There is no continuous chiral symmetry behind this relation, but
only a discrete one $\eta\to \gamma_5\rho_2 \eta$. For further fermionic
symmetries see the Appendix.
The argument presented in \cite{Lucini:2015hfa}
for the case of the $C$-boundary condition 
also applies for the $CP$-boundary:
for those symmetries that are broken
by the boundary, the breaking is expected to fall off exponentially in $T$.

Another solution to the complex determinant problem can be given
for two degenerate flavors. Then the recipe is to add an extra
rotation in flavor space at the boundary as
\begin{align}
    \label{eq:eq1}
    \eta(x,y,z,t+T) &= -\gamma_5\gamma_x\rho_2\tau_1\ \eta(L-x,y,z,t),
\end{align}
where $\tau_i$'s are Pauli-matrices in flavor space, and now the $\eta$ carries
a flavor index, too.
This makes the fermion
path integral real, since the Dirac-operator is $\gamma_5\tau_3$-Hermitian.
This solution can be written in the usual four-component spinor basis, and it can be
implemented in a similar way as the so-called $G$-parity boundary condition \cite{Bai:2015nea}.
Equation \ref{eq:eq1} can be also applied to unrooted staggered fermions.

To get the expression for the two point pion correlation function, the interpolating operators are needed in the Majorana basis:
\begin{align}
\begin{split}
\mathcal{O}_{\pi^{-}} &= \overline{\psi}_u \gamma_5 \psi_d = -\frac{1}{2}
\eta_u^{T} \gamma_5 C \left( 1 - \rho_2 \right) \eta_d\\
\overline{\mathcal{O}}_{\pi^{-}} &= - \overline{\psi}_d \gamma_5 \psi_u =
\frac{1}{2} \eta_d^{T} \gamma_5 C \left( 1 - \rho_2 \right) \eta_u.
\end{split}
\end{align}
The correlator between $x$ and $y$ then is
\begin{widetext}
\begin{align}
\left\langle \mathcal{O}_{\pi^{-}}(x) \, \overline{\mathcal{O}}_{\pi^{-}}(y)
\right\rangle = - \frac{1}{4} \left\langle \big(\eta_u^{T})_x \gamma_5
  C \left( 1 - \rho_2 \right) \big(\eta_d\big)_x \big(\eta_d^{T}\big)_y
\gamma_5 C \left( 1 - \rho_2 \right) \big(\eta_u\big)_y \right\rangle,
\end{align}
which after integrating out the Grassmann fields and choosing
$m_u=m_d$ becomes
\begin{align}
\left\langle \mathcal{O}_{\pi^{-}}(x) \, \overline{\mathcal{O}}_{\pi^{-}}(y)
\right\rangle = \frac{1}{4} \mathsf{Tr} \left[ \big(\hat{D}^{-1}\big)_{y,x}\,
  \rho_2\,  \big(\hat{D}^{-1}\big)^{\dagger}_{x,y}\, \rho_2 +
  \big(\hat{D}^{-1}\big)_{y,x}\,   \big(\hat{D}^{-1}\big)^{\dagger}_{x,y} \right].
\end{align}
\end{widetext}
More point correlations functions can be constructed analogously.

\begin{figure}
    \centering
    \includegraphics{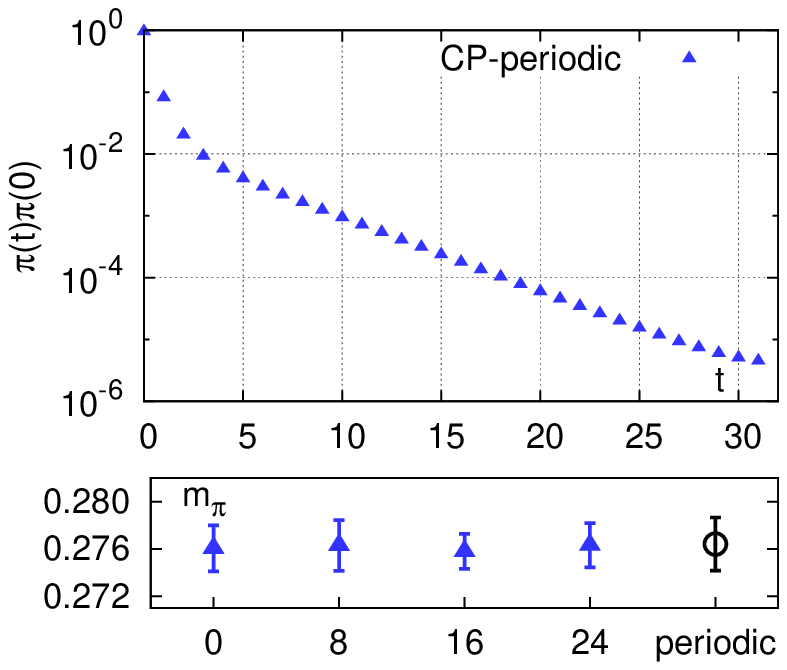}
    \caption{
	\label{fi:pion}
	{\it Upper:} Pion propagator with $CP$-boundary condition.
	{\it Lower:} Pion masses obtained after shifting the gauge field by 0, 8, 16, and 24 units in t-direction (filled), and
	with periodic boundary condition (open).
    } 
\end{figure}

As a first exploratory study we implemented the $CP$-boundary condition with a
Wilson-Dirac operator. Quenched configurations were generated with
$CP$-boundary and we found, that the observables and in particular their
autocorrelation times were consistent with the respective $P$-boundary condition values.
To study the above proposal for fermions, we took $CP$-boundary configurations
generated at $\beta=4.35$ on $16^3\times32$ lattices, $w_0=1.57$.
Four steps of
stout smearing with smearing parameter $0.125$ were applied \cite{Morningstar:2003gk}.
The $\pi^+$ pion propagator
was measured at the bare Wilson-mass $-0.16$, see upper panel of Figure
\ref{fi:pion}. Contrary to the usual periodic case the
propagator is a single exponential for times far away from $t=0$ and $T$. There is no backward contribution
coming from the anti-particle $\pi^-$.
The reason for this is, that the anti-particle $\pi^-$ is now
transformed under $CP$ conjugation to a $-\pi^+$. The corresponding propagation amplitude will be zero,
since the matrix element of the pion annihilating operator between the $\pi^+$ and vacuum states is zero.
To demonstrate the translational invariance,
we also show fitted pion masses, that were obtained after shifting the gauge
configuration in $t$-direction by 0, 8, 16, and 24 slices. The mass values are
compatible with each other and also with the mass obtained from a lattice of same size with
periodic boundary condition.

\section*{Discussion and Outlook}

The last sections show that a lot of properties of observables
in the $P$-periodic setting are similar to or better than those in the open
setting where the field space is connected. In particular the autocorrelation
time of local quantities like the topological charge on a time slice improves
to the level of open boundary conditions without showing boundary artefacts.
But this does not mean that the
field space in the $P$-periodic setup is connected. This can be seen for example in the
instanton picture. In the periodic case the topological sectors are labeled by
the net number of instantons with positive and negative charge $N_+-N_-$, which
is an integer and conserved. The total instanton number $N_++N_-$ is only
conserved $\pmod{2}$ due to instanton pair creation and annihilation. For
open lattices $N_+$, $N_-$, and $N_+\pm N_-$ are not well defined, so
instantons do not lead to multiple sectors in field space. For the $P$-periodic
case instantons can propagate once around the $P$-periodic direction and become
anti-instantons, so $N_+$, $N_-$, and $N_+-N_-$ are also not well defined. But
the total instanton number $N_++N_-$ is conserved by this process and like on
the torus well defined $\pmod{2}$. This suggests there are still two
sectors in field space which correspond to the even and odd sectors on the
torus, and that the same lattice artefacts are necessary to move between
these sectors. Then also the critical slowing down affects the
tunneling rate identically as on the torus. But in the case with $P$-periodic boundary
conditions, i.e. only two sectors, it is much simpler to simulate long
enough such that all sectors are sampled well. Already this makes the simulation at smaller
lattice spacings feasible.

Having only two sectors also opens up possibilities to study even
smaller lattice spacings with completely frozen topology if the relative
weight of the two sectors is known. This can be achieved for example by
integrating up the relative weight of the two sectors starting at a
coupling where there is still enough tunneling between the sectors. This
strategy has recently been applied at high temperatures where there are
only two sectors relevant as well.\cite{Borsanyi:2016ksw,Frison:2016vuc}
$P$-periodic boundaries make this method applicable also at low
temperatures where on the torus at reasonable volumes many sectors
with different relative weights contribute.

In this article we have shown how to formulate and simulate lattice QCD on a
non-orientable space-time manifold. For the pure gauge case our simulations show that
this change in the topology of space-time leads to a strong reduction of the
autocorrelation time of the topological charge density comparable to the improvement
observed using open boundary conditions. While the pure gauge case is straightforward
to construct and its implementation is virtually free of additional numerical cost,
the inclusion of fermions is not trivial. We have demonstrated how to combine
non-orientable space-time with charge conjugation or flavor symmetry to get a real
fermionic contribution to the action and also how to measure fermionic observables.
Testing the new boundaries for dynamical fermions is left for future work. It
will be especially interesting to compare the numerical costs to that of eg.
open boundary conditions. Since it depends on the implementation details of
fermions in the Majorana basis, we refrain from a precise cost estimation at this
point.

\section*{Acknowledgments}
The authors thank Tamas Kovacs and Daniel Nogradi for discussions and useful comments on the manuscript.
S.M. thanks Jakob Simeth, Florian Rappl, and Rudolf
R\"odl for many discussions on possible applications of non-orientable manifolds
in particle physics. The simulations were performed on the GCS supercomputers
JURECA and JUQUEEN in JSC in J\"ulich, on SuperMUC at LRZ in Garching, and on
the QPACE machines funded by the DFG grant SFB/TR55.

\appendix

\section*{Appendix}
\label{sec:appendix}

In this Appendix we list the symmetries of the single flavor fermion action
in the Majorana representation, i.e. in the basis of eigenstates of the
charge conjugation operation. For the definition see \cite{Lucini:2015hfa}.
We use the following representation
for the charge conjugation matrix:
$C=i\gamma_y\gamma_t=C^\dagger=C^{-1}=-C^t$. The usual fermion action is
\begin{gather*}
    S_f[\psi,\overline{\psi},U]= \overline{\psi} D[U] \psi.
\end{gather*}
Applying the following transformation 
\begin{gather*}
    \eta=
    \frac{1}{\sqrt{2}}
    \begin{pmatrix}
	1 & C \\
	-i & iC
    \end{pmatrix}
    \begin{pmatrix}
	\psi \\
	\overline{\psi}^t
    \end{pmatrix},
\end{gather*}
we get 
\begin{gather*}
    S_f[\eta,U]= -\frac{1}{2}\eta^t C \hat{D}[U] \eta,
\end{gather*}
where $\hat{D}[U]= D[\mathrm{Re}U\cdot 1_{2\times 2} -i\mathrm{Im}U\cdot \rho_2]$.

\subsection{Infinite volume}

First we start with the symmetries of the fermion action in infinite volume.
The symmetries also apply in a finite box with periodic boundary conditions.
The transformation for the gauge fields is standard and not shown explicitly.

\begin{enumerate}

    \item U(1) phase transformation:
	\begin{gather*}
	    \psi \to \exp(i\theta) \psi \quad\text{ and }\quad \overline{\psi}\to \overline{\psi} \exp(-i\theta),\\
	    \eta\to \exp(-i\theta \rho_2) \eta.
	\end{gather*}

    \item U(1) chiral transformation for a massless fermion in the continuum:
	\begin{gather*}
	    \psi \to \exp(i\theta\gamma_5) \psi \quad\text{ and }\quad \overline{\psi}\to \overline{\psi} \exp(i\theta\gamma_5),\\
	    \eta\to \exp(i\theta \gamma_5) \eta.
	\end{gather*}

    \item Charge conjugation:
	\begin{gather*}
	    \psi \to C\overline{\psi}^t \quad\text{ and }\quad \overline{\psi}\to -\psi^t C,\\
	    \eta\to \rho_3\eta.
	\end{gather*}

    \item Translations along the $\mu$-axis with $x'_\mu = x_\mu+1$:
	\begin{gather*}
	    \psi(x) \to \psi(x') \quad\text{ and }\quad \overline{\psi}(x) \to \overline{\psi}(x'),\\
	    \eta(x) \to \eta(x').
	\end{gather*}

    \item $\mu$-axis reversal with $x'_\mu= -x_\mu$:
	\begin{gather*}
	    \psi(x) \to \gamma_5\gamma_\mu\psi(x') \quad\text{ and }\quad \overline{\psi}(x)\to \overline{\psi}(x')\gamma_\mu\gamma_5,\\
	    \eta(x) \to -\gamma_5\gamma_\mu\rho_2\eta(x').
	\end{gather*}
\begin{widetext}
    \item 90$^{\circ}$-rotation in $\mu\nu$-plane with $x'_\mu=-x_\nu, x'_\nu=x_\mu$:
        \begin{gather*}
            \psi(x) \to \exp \!\left(-i \frac{\pi}{2} \sigma_{\mu\nu} \right) \psi(x') 
            \quad\text{ and }\quad \overline{\psi}(x)\to \overline{\psi}(x')\exp \!\left(i \frac{\pi}{2} \sigma_{\mu\nu} \right),\\
            \eta(x) \to  \exp \!\left(-i \frac{\pi}{2} \sigma_{\mu\nu} \right) \eta(x'),
        \end{gather*}
        where $\sigma_{\mu\nu} = -\frac{i}{4} \left[ \gamma_\mu, \gamma_\nu \right]$.
\end{widetext}
\end{enumerate}

\subsection{$CP$-boundary}
Here we list the symmetries of the theory with $CP$-boundary conditions. In the spatial
directions we have usual periodic boundary conditions, whereas in the time direction we have
\begin{align*}
    U_{x}(x,y,z,t+T)&= U^t_{x}(L-x-1,y,z,t),\\
    U_{i}(x,y,z,t+T)&= U^{*}_{i}(L-x,y,z,t), \quad i=y,z,t
\end{align*}
for the gauge fields and
\begin{gather*}
    \eta(x,y,z,t+T) = -\gamma_5\gamma_x\rho_2\rho_3\ \eta(L-x,y,z,t)
\end{gather*}
for the fermions. The symmetries are:

\begin{enumerate}

    \item Sign transformation:
	\begin{gather*}
	    \psi \to -\psi \quad\text{ and }\quad \overline{\psi} \to -\overline{\psi},\\
	    \eta \to -\eta.
	\end{gather*}

    \item Combined chiral and $U(1)$ phase transformation for a massless fermion in the continuum:
	\begin{gather*}
	    \psi \to -\gamma_5 \psi \quad\text{ and }\quad \overline{\psi} \to \overline{\psi}\gamma_5,\\
	    \eta \to \gamma_5\rho_2 \eta.
	\end{gather*}

    \item Charge conjugation combined with a $U(1)$ phase transformation:
        \begin{gather*}
            \psi \to i C\overline{\psi}^t \quad\text{ and }\quad \overline{\psi}\to i \psi^t C,\\
            \eta\to -i \rho_2 \rho_3\eta.
        \end{gather*}

    \item Translations along the $\mu$-axis for $\mu=y,z,t$.

    \item $\mu$-axis reversal for $\mu=y,z,t$.

    $\mu$-axis reversal combined with a $U(1)$ phase transformation for $\mu=x$:
	\begin{gather*}
	    \psi(x) \to i\gamma_5\gamma_\mu\psi(x') \quad\text{ and }\quad \overline{\psi}(x)\to -i\overline{\psi}(x')\gamma_\mu\gamma_5,\\
	    \eta(x) \to i\gamma_5\gamma_\mu \eta(x').
	\end{gather*}

    \item 90$^{\circ}$-rotation in the $yz$ plane.

\end{enumerate}


%

\end{document}